\documentclass[12pt]{article}
\usepackage{amssymb}
\usepackage[dvips]{epsfig}

\newcommand{\be}{\begin{equation}}
\newcommand{\ee}{\end{equation}}
\def\bea{\begin{eqnarray}}
\def\eea{\end{eqnarray}}

\newcommand{\bn}{\begin{eqnarray}}
\newcommand{\en}{\end{eqnarray}}

\newcommand{\tc}{\tilde{C} }

\newcommand{\p}{\partial}

\newcommand{\nn}{\nonumber}

\newcommand{\no}{\noindent}

\newcommand{\tb}{\tilde{B}}
\newcommand{\ta}{\tilde{A}}

\newcommand{\s}{\,\,\,\,}
\def\bea{\begin{eqnarray}}
\def\eea{\end{eqnarray}}

\newcommand{\beq}{\begin{eqnarray}}
\newcommand{\eeq}{\end{eqnarray}}
\begin{document}

\title{\textbf{Massive spin-2 particle from a rank-2 tensor}}
\author{D. Dalmazi\footnote{dalmazi@feg.unesp.br} \\
\textit{{UNESP - Campus de Guaratinguet\'a - DFQ} }\\
\textit{{Avenida Dr. Ariberto Pereira da Cunha, 333} }\\
\textit{{CEP 12516-410 - Guaratinguet\'a - SP - Brazil.} }\\}
\date{\today}
\maketitle

\begin{abstract}

Here we obtain all possible second-order theories for a rank-2
tensor which describe a massive spin-2 particle. We start with a
general second-order Lagrangian with ten real parameters. The
absence of lower spin modes and the existence of two local field
redefinitions leads us to only one free parameter. The solutions
split into three  one-parameter classes according to the local
symmetries of the massless limit. In the class which contains the
usual massive Fierz-Pauli theory, the subset of spin-1 massless
symmetries is maximal. In another class where the subset of spin-0
symmetries is maximal, the massless theory is invariant under Weyl
transformations and the mass term does not need to fit in the form
of the Fierz-Pauli mass term. In the remaining third class neither
the spin-1 nor the spin-0 symmetry is maximal and we have a new
family of spin-2 massive theories.

\end{abstract}

\newpage

\section{Introduction}

In general relativity the gravitational interaction is mediated by an apparently massless spin-2 particle. In
order to understand whether the graviton is really massless one investigates the consequences of turning on a
(tiny) mass term. The mass discontinuity \cite{vdv,zak} and the appearance of ghosts \cite{db} are two known
longstanding problems of the massive theory. More recently, motivated also by experimental large scale
gravitational results, there has been an intense work on massive gravity where the mass discontinuity problem
has been addressed via the ideas of \cite{vain}, see for instance \cite{rgt,rg} and \cite{hr}. See also the
review works \cite{rt08,h11}. Besides ghosts and mass discontinuity, there is also an ongoing discussion
\cite{dgnr,gru,rgt2,dw,frhm} on causality in massive gravity theories.

We would like to stress that all the above works are based on the usual description of massive spin-2 particles
suggested by Fierz and Pauli (FP) long ago \cite{fp} where the basic field is a symmetric rank-2 tensor. It is
certainly welcome to search for alternative descriptions of massive spin-2 particles. Among them, probably the
most natural, specially if we have in mind a frame-like ($e_{\mu}^{\,\, a}$) description of gravity, is to allow
for an arbitrary rank-2 tensor with non vanishing antisymmetric part $e_{[\mu \, a]} \ne 0$. This is the route
we follow here and has been followed before in, e.g., \cite{rivers,barnes,bw,pvn73,cko}.

The conclusion of those works is that the only possibility, in the
massive case, is the well known symmetric description of FP.
Regarding the massless spin-2 case, although  \cite{pvn73}
concludes in favor of the linearized Einstein-Hilbert theory
(massless FP ) as the only possibility, one has found in
\cite{abgv} another theory which is invariant under transverse
linearized reparametrizations and Weyl transformations and
correctly describes a massless spin-2 particle in terms of a
symmetric tensor. Moreover in \cite{cmu} there is a further
description of a massless spin-2 particle in terms of an arbitrary
rank-2 tensor. Back to the massive case, there are two exceptions
to the conclusion of \cite{rivers}-\cite{cko} recently found in
\cite{ms} and \cite{nfp}. In both \cite{ms,nfp}, $e_{[\mu\nu]}$
does not decouple from the symmetric part $e_{(\mu\nu)}$ in any
local way at action level. In particular, in \cite{nfp} the mass
term $-m^2(e_{\mu\nu}e^{\nu\mu} + c\, e^2 )$ does not need to fit
in the usual FP form with $c=-1$. The real parameter $c$ is
arbitrary.

The above exceptions have prompted us to revisit the works \cite{rivers}-\cite{cko} in order to achieve a
complete classification of massive spin-2 particles in terms of an arbitrary rank-2 tensor. In section 2 we
start with a general second-order Lagrangian density with ten free real parameters. By requiring the existence
of only one massive spin-2 pole in the propagator and using a trivial field redefinition (\ref{r}) we get rid of
eight parameters, see (\ref{pmenos}),(\ref{pmais}),(\ref{dmais}),(\ref{c5}),(\ref{c6}),(\ref{c7}),(\ref{a2a3}).
There is another trivial field redefinition (\ref{w}) which allows us to fix another undetermined parameter in
section 3 where we also analyze the local symmetries of the corresponding massless theories. Based on those
symmetries we end up with three classes of models, one of them is new. Its equations of motion are presented in
section 4 where we also discuss the mass discontinuity problem. In section 5 we draw our conclusions.

\section{General setup}

We start with a second-order Lagrangian density in $D=4$ with ten real free parameters\footnote{Throughout this
work we use $\eta_{\mu\nu}=diag(-,+,+,+)$ and the abbreviations LEH (linearized Einstein-Hilbert) and GR
(general relativity) among others defined in the text.}

\bea {\cal L}_G\left\lbrack e_{\mu\nu} \right\rbrack &=& a_1
\left(\p^{\alpha}e_{\alpha\beta}\right)^2 + a_2
\left(\p^{\alpha}e_{\alpha\nu}\right)\left(\p_{\beta}e^{\nu\beta}\right)
+ a_3 \left(\p^{\nu}e_{\mu\nu}\right)^2 + b_1 e \, \Box \, e + b_2
\, \p^{\mu}e\, \p^{\alpha}e_{\alpha\mu} \nn \\ &+& \frac{p_1}2
e_{\mu\nu}\Box e^{\mu\nu} + \frac{p_2}2 e_{\mu\nu}\Box e^{\nu\mu}
+ c\, e^2 + d_1 \, e_{\mu\nu}e^{\mu\nu} + d_2 \,
e_{\mu\nu}e^{\nu\mu} \quad . \label{lg} \eea

\no It is convenient to rewrite ${\cal L}_G$ in terms of
antisymmetric ($B_{\mu\nu}$) and symmetric ($h_{\mu\nu}$) tensors.
Using $e_{\mu\nu} = B_{\mu\nu} + h_{\mu\nu}$ we obtain

\bea {\cal L}_G\left\lbrack B_{\mu\nu}, h_{\mu\nu} \right\rbrack &=& S \left(\p^{\alpha}h_{\alpha\beta}\right)^2
 + b_1 h \, \Box \, h  + b_2 \, \p^{\mu}h \, \p^{\alpha}h_{\alpha\mu} + c\, h^2 \nn \\ &+& p_+ \, h_{\mu\nu}\Box
 h^{\mu\nu} + d_+ h_{\mu\nu}^2 + 2 (a_1-a_3)\p^{\mu}B_{\mu\nu}\, \p_{\alpha}h^{\alpha\nu} \nn\\ &+&
 (a_1 + a_3 - a_2)\p^{\mu}B_{\mu\nu}\p_{\alpha}B^{\alpha\nu} + p_- \, B_{\mu\nu}\Box B^{\mu\nu} +
 d_- \, B_{\mu\nu}B^{\mu\nu} \label{lgbh} \eea

\no where

\bea S &=& a_1 + a_2 + a_3 \quad , \label{s} \\  d_{\pm} &=& d_1
\pm d_2 \quad . \label{dpm} \\
 p_{\pm} &=& (p_1
\pm p_2)/2 \quad . \label{ppm} \eea

Our aim is to single out the regions in the 10-dimensional parameters space of (\ref{lg}) which correspond to
only one massive spin-2 particle without tachyons and ghosts. We assume that all parameters in the second
derivative terms are dimensionless and in the massless limit all non-derivative terms vanish, i.e.,

\be \lim_{m \to 0} \left(c,d_1,d_2\right) = (0,0,0) \quad . \label{ml} \ee

As far as we known there are three examples of massive spin-2 models in the literature \cite{fp,ms,nfp} in
agreement with our hypothesis. For all three models we have

\be (a_2,a_3)=(1/2,1/4) \quad , \quad p_1=p_2=1/2 \quad , \quad (d_1,d_2) = (0,-m^2/2) \quad . \label{todos}\ee

 The first model is the well known massive Fierz-Pauli (FP) theory  \cite{fp} which can be entirely described
by means of a symmetric tensor. Its massless limit is the linearized Einstein-Hilbert theory. It is invariant
under linearized reparametrizations $\delta e_{\mu\nu} = \p_{\mu}\xi_{\nu} + \p_{\nu}\xi_{\mu} $. The massive FP
theory corresponds to

\be \left(a_1,b_1,b_2,c \right) = \left(1/4, -1/2,-1,m^2/2 \right) \quad , \quad S=1 \quad . \label{fp}\ee

\no Since $B_{\mu\nu}$ only appears now in the last term of (\ref{lgbh}), the parameter $d_-$ is a free
parameter in the FP model.

 In the second model, defined in \cite{ms}, there is a nontrivial coupling between the symmetric ($h_{\mu\nu}$) and
antisymmetric ($B_{\mu\nu}$) parts of the rank-2 tensor. The massless limit of \cite{ms} is invariant under
$\delta e_{\mu\nu} = \p_{\mu}\xi_{\nu} $. For the massive model \cite{ms} we have

\be \left(a_1,b_1,b_2,c \right) = \left(-1/4, 0 , 0 ,m^2/2 \right) \quad , \quad S=1/2 \quad . \label{ms}\ee

Regarding the third model defined in \cite{nfp}, the massless limit, previously studied in \cite{cmu}, is
invariant under $\delta e_{\mu\nu} = \p_{\mu}\xi_{\nu} + \eta_{\mu\nu} \Lambda $. While the mass term of the
models \cite{fp} and \cite{ms} must be of the Fierz-Pauli type $\left(e_{\mu\nu}e^{\nu\mu} - e^2\right)$, in the
massive model \cite{nfp} one can add a term proportional to the square of the trace of the rank-2 tensor $\left(
e_{\mu\nu}e^{\nu\mu}/2 + f e^2\right)$ where $f$ is an arbitrary real constant. This is a consequence of the
linearized Weyl symmetry of the massless theory. This symmetry can be extended to the whole massive theory if we
choose $f=-1/4 $ which  allows the use of a traceless rank-2 tensor. The massive model of \cite{nfp} is
described in general by

\be \left(a_1,b_1,b_2 \right) = \left(-1/12, -1/6 , -1/3 \right) \quad , \quad S=2/3 \quad . \label{nfp}\ee

\no while the parameter $c$ remains arbitrary for reasons already mentioned.

The Lagrangian ${\cal L}_G$ can be written as ${\cal L}_G = e_{\mu\nu}G^{\mu\nu\alpha\beta}e_{\alpha\beta}$
where, suppressing the indices, we have the differential operator

\be G = \left(d_+ + p_+\Box \right)P_{TT}^{(2)} + \left(d_- + p_-\Box \right)P_{AA}^{(0)} +
\sum_{s=0,1}\sum_{I,J}A_{IJ}^{(s)}P_{IJ}^{(s)} \quad , \label{g} \ee

\no where the spin-s operators $P_{IJ}^{(s)}$, given in the appendix A, satisfy the algebra

\be P_{IJ}^{(s)} P_{KL}^{(r)} = \delta^{rs}\delta_{JK} P_{IL}^{(s)} \quad . \label{algebra} \ee

\no The operators $P_{IJ}^{(1)}$ with $I,J = S,A$ form a subalgebra of (\ref{algebra}) as well as the operators
$P_{IJ}^{(0)}$ with $I,J = T,W$. The $2 \times 2$ matrices $A_{IJ}^{(s)}$ are given by:

\bea A_{AA}^{(1)} &=& d_+ + \left( p_+ -  S/2 \right)\Box \label{ass} \\
A_{SS}^{(1)} &=&  d_- + \left(2 p_- + a_2 - a_1 - a_3 \right) \frac{\Box}2 \label{aaa} \\
A_{AS}^{(1)} &=& A_{SA}^{(1)} =  (a_1-a_3)\Box/2 \label{aas}\\
 A_{TT}^{(0)} &=& d_+ + 3\, c + \left(p_+ + 3 \, b_1 \right)\Box \label{att} \\
A_{WW}^{(0)} &=& d_+ + c + \left(p_+ + b_1 - S-b_2 \right)\Box \label{aww} \\
A_{TW}^{(0)} &=& A_{WT}^{(0)} = \sqrt{3}\left\lbrack c + \left(b_1
- b_2/2\right)\Box \right\rbrack \label{atw} \eea

\no A key role will be played by the propagator which is proportional to the operator
$G^{-1}_{\mu\nu\alpha\beta}$ which is given, suppressing the indices again, by

\be G^{-1} = \frac{P_{TT}^{(2)}}{d_+ + p_+\Box} + \frac{P_{AA}^{(0)}}{d_- + p_-\Box } +
\sum_{s=0,1}\sum_{I,J}\left(A^{-1}\right)_{IJ}^{(s)}P_{IJ}^{(s)}  \label{gmenos1} \ee

\no where the inverse matrix $\left( A^{-1}\right)_{IJ}^{(s)}$ is given explicitly by

\be \left(A^{-1}\right)_{11}^{(s)} = \frac{A_{22}^{(s)}}{K^{(s)}} \quad ; \quad  \left(A^{-1}\right)_{22}^{(s)}
= \frac{A_{11}^{(s)}}{K^{(s)}} \quad ; \quad \left(A^{-1}\right)_{12}^{(s)} = \left(A^{-1}\right)_{21}^{(s)} =
-\frac{A_{12}^{(s)}}{K^{(s)}} \quad . \label{inverse} \ee

\no with the determinants:

\be K^{(s)} = A_{11}^{(s)} A_{22}^{(s)}- \left\lbrack A_{12}^{(s)}\right\rbrack^2  \quad , \quad s=0,1 \quad .
\label{ksg} \ee

In $G^{-1}$ we have four sources of poles, namely, the operators in the denominators below  $P_{TT}^{(2)}$ and
$P_{AA}^{(0)}$ and the determinants $K^{(s)}$ for $s=0,1$. We must determine the parameters of our model such
that we only have one massive physical pole coming from the denominator below $P_{TT}^{(2)}$. The no-pole
condition in the denominator of $P_{AA}^{(0)}$ requires :

\be p_1 - p_2 = 0 \quad . \label{pmenos} \ee

\no In the operator below $P_{TT}^{(2)}$ we have a massive particle with $m^2=-d_+/p_+$. From the two-point
amplitude saturated with external sources $T_{\mu\nu}$ we can calculate the residue at $k^2 \to - m^2$, whose
imaginary part is given in momentum space by

\bea I_{-m^2} &=& \Im \lim_{k^2 \to - m^2}(k^2 + m^2)(- \frac i2 )
T_{\mu\nu}^*(k)\left\lbrack G^{-1}(k,-k)
\right\rbrack^{\mu\nu\alpha\beta} T_{\alpha\beta}(k) \nn \\ &=&
\frac 1{p_+} T_{\mu\nu}^*(k)\left\lbrack P_{TT}^{(2)}\right\rbrack
^{\mu\nu\alpha\beta} T_{\alpha\beta}(k) \equiv \frac 1{p_+} T^*
P_{TT}^{(2)} T \label{im} \eea

\no Since, see for instance \cite{nfp}, $T^* P_{TT}^{(2)} T > 0 $ at $k^2 = - m^2$, a physical particle
($I_{-m^2} > 0 $) requires $p_+ > 0$. After a dilatation $e_{\mu\nu} \to \Lambda e_{\mu\nu}$ we can set without
loss of generality

\be p_+ = p_1 = p_2 = 1/2 \quad , \label{pmais} \ee

\no and the mass is fixed by $d_+$

\be d_+ = - m^2/2 \quad . \label{dmais} \ee

\no Let us now examine the consequences of the no-pole condition
on the determinants $K^{(s)}$. From (\ref{ksg}) and
(\ref{ass})-(\ref{atw}) we can write down

\be K^{(s)} = C_2^{(s)}\Box^2 + C_1^{(s)}\Box +  C_0^{(s)} \quad . \label{ks} \ee

\no The existence of $G^{-1}$ and the absence of poles in $K^{(s)}$ require

\bea C_2^{(1)} &=& \left(a_1 - 1/4\right)\left(a_3 - 1/4\right) - \left(a_2 - 1/2 \right)^2 =0
\quad , \label{c4} \\
C_1^{(1)} &=& d_-  \left( 1 - S \right)/2 + m^2\left(a_2-a_1-a_3\right)/4 = 0 \quad, \label{c5} \\
C_0^{(1)} &=& - m^2 d_-/2 \ne 0 \quad , \label{d1} \\
C_2^{(0)} &=& 3 \left( b_1 + 1/6 \right)\left( 2/3 - S\right) - 3\left(b_2 + 1/3
\right)^2/4 = 0 \quad , \label{c6} \\
C_1^{(0)} &=& m^2 \left(S-1+b_2 - 4\, b_1\right)/2 + 3\, c\left(2/3 - S \right) = 0 \quad, \label{c7}
\\ C_0^{(0)} &=& m^2\left(m^2/8 - c\right) \ne 0 \quad , \label{d2} \eea

\no Henceforth we assume $d_- \ne 0$ and $c\ne m^2/8$. The seven conditions
(\ref{pmenos}),(\ref{pmais}),(\ref{dmais}),(\ref{c4}),(\ref{c5}),(\ref{c6}),(\ref{c7}) still leave 3 out of 10
coefficients in (\ref{lg}) arbitrary. Part of this redundancy is due to the following two families of local
field redefinitions:

\be e_{\mu\nu} \to e_{\mu\nu} + \frac a2 \eta_{\mu\nu}\, e \quad ;
\quad a \ne - 1/2 \quad ,  \label{w} \ee

\be e_{\mu\nu} \to A\, e_{\mu\nu} + (1-A) e_{\nu\mu} \quad ; \quad A \ne 1/2 \quad , \label{r} \ee

\no The restrictions $ a \ne - 1/2$ and $A \ne 1/2 $ are needed
for the existence of the inverse transformations. The
transformations (\ref{r}) are dilations in the antisymmetric part
of the tensor, i.e., $(h_{\mu\nu},B_{\mu\nu}) \to  (h_{\mu\nu},
(2A-1)B_{\mu\nu})$.

From (\ref{w}) and (\ref{r}) we can fix two more coefficients such that all solutions to the no-pole conditions
(\ref{c4}),(\ref{c5}),(\ref{c6}),(\ref{c7}) become one-parameter families. More specifically, the
transformations (\ref{w}) and (\ref{r}) lead to

\be b_2 \to b_2 + a \left( S + 2 b_2 \right) \quad , \label{w1} \ee

\be c \to c + 4\, a (a+1) \left( c - m^2/8 \right) \quad ,
\label{w2} \ee

\be b_1 \to b_1 + a \left\lbrack (a+1)(4\, b_1 + 1/2 )-  a/4 S -
(a+1/2)b_2 \right\rbrack \quad , \label{w3} \ee

\be a_2 \to a_2 + 2\, A(1-A)(a_1 + a_3 - a_2) \quad , \label{r1} \ee

\be a_1 \to  A^2(a_1 + a_3 - a_2) + A(a_2-2\, a_3) + a_3 \quad , \label{r2} \ee

\be a_3 \to  A^2(a_1 + a_3 - a_2) + A(a_2-2\, a_1) + a_1 \quad , \label{r3} \ee

\be d_- \to d_-(1-2\, A)^2 \quad ; \quad p_- \to p_- (1-2\, A)^2 \quad , \label{r4} \ee

\no while $d_+$, $p_+$ and the constraints (\ref{c4}),(\ref{c5}),(\ref{c6}),(\ref{c7}) are invariant. The sum
$S=a_1+a_2+a_3$ is also invariant under (\ref{w1})-(\ref{r4}), it will play an important role in the
classification of the solutions to (\ref{c4}),(\ref{c5}),(\ref{c6}) and (\ref{c7}).

In the appendix we prove that due to (\ref{r}) we can choose without loss of generality the following solution
to (\ref{c4})

\be a_2 = 1/2 \quad ; \quad a_3 = 1/4 \quad . \label{a2a3} \ee

\no The solution (\ref{a2a3}) is assumed henceforth unless otherwise stated. Due to the special role, see
(\ref{w1}), of the fixed point
 $S=-2 b_2$ we postpone fixing $b_2$ by using (\ref{w}).

In summary, any one parameter family of solutions to (\ref{c5}),(\ref{c6}) and (\ref{c7}) with (\ref{pmenos}),
(\ref{pmais}),(\ref{dmais}),(\ref{d1}),(\ref{d2}) and (\ref{a2a3}) describes one massive physical particle of
spin-2. In the next section we split those solutions into three classes.

\section{Classifying the massive models via the massless limit}

As we will see in this section, the massless limit, defined in (\ref{ml}), of the  massive spin-2 models which
satisfy the constraints (\ref{c4}) and (\ref{c6}) necessarily have local symmetries which can help us
classifying the solutions to the constraints (\ref{c5}),(\ref{c6}) and (\ref{c7}).

First, it is clear from (\ref{g}) and the algebra (\ref{algebra}) that there can only be spin-1 and spin-0 local
symmetries of (\ref{lg}). Moreover, since each of the sets of operators $\left\lbrace P_{AA}^{(0)}\right\rbrace
$,$ \left\lbrace P_{IJ}^{(1)}\right\rbrace $ and $\left\lbrace P_{KL}^{(0)}\right\rbrace$, with $(K,L) \ne
(A,A)$ , form subalgebras of the algebra (\ref{algebra}), we can write down the most general symmetry
transformation of $e_{\mu\nu}$ as follows,

\be \delta e_{\mu\nu} = \delta^{(0)}_{AA}e_{\mu\nu} + \delta^{(0)}e_{\mu\nu} +  \delta^{(1)}e_{\mu\nu} \quad .
\label{deltaeg} \ee

\no where

\bea \delta^{(0)}_{AA}e_{\mu\nu} &=& E\left\lbrack P_{AA}^{(0)}\right\rbrack_{\mu\nu\alpha\beta}
\Lambda^{\alpha\beta} \quad , \label{deltaaa} \\
\delta^{(0)}e_{\mu\nu} &=&  \left\lbrack A\, P_{TT}^{(0)} + B
P_{WW}^{(0)} + C \sqrt{3} \left( P_{TW}^{(0)} +
P_{WT}^{(0)}\right) \right\rbrack_{\mu\nu\alpha\beta} \Lambda^{\alpha\beta} \quad , \label{delta0} \\
\delta^{(1)}e_{\mu\nu} &=&  \left\lbrack \tilde{A}\, P_{SS}^{(1)}
+ \tilde{B} P_{AA}^{(1)} + \tilde{C} \left( P_{AS}^{(1)} +
P_{SA}^{(1)}\right) \right\rbrack_{\mu\nu\alpha\beta}
\Lambda^{\alpha\beta} \quad . \label{delta1}  \eea

\no The parameter $\Lambda^{\alpha\beta}(x)$ is an arbitrary rank-2 tensor while $A,B,C,\ta,\tb,\tc,E$ are
arbitrary real constants. Due to the fact that $P_{IJ}^{(s)}\Lambda $ is independent of $P_{KL}^{(r)}\Lambda $
unless $(I,J,s)=(K,L,r)$ we have from $\delta S_{G} = 2 \int d^4x e_{\alpha\beta}G^{\alpha\beta\mu\nu}\delta
e_{\mu\nu}=0$, with $(a_2,a_3)$ arbitrary,  the following set of equations:

\bea \left(b_1 + 1/6 \right)\, A + \left( b_1 - b_2/2 \right)
 C &=& 0 \quad , \label{s01} \\
\left( b_1 - b_2/2 \right) A +
\left( b_1 - b_2 - S + 1/2  \right)\, C &=& 0 \quad , \label{s02} \\
\left( b_1 - b_2 - S + 1/2  \right)\, B + 3 \left( b_1 - b_2/2 \right)\, C &=& 0 \quad , \label{s03} \\
\left( b_1 - b_2/2 \right)\, B + 3\left(b_1 + 1/6
\right)\, C &=& 0  \quad , \label{s04} \\
(1-S)\ta + (a_1-a_3)\tc &=& 0 \quad , \label{s11} \\
(a_1-a_3)\ta + (a_2-a_1 - a_3)\tc &=& 0 \quad , \label{s12} \\
(a_2-a_1 - a_3)\tb + (a_1-a_3)\tc &=& 0 \quad , \label{s13} \\
(a_1-a_3)\tb + (1-S)\tc &=& 0 \quad , \label{s14} \\
d_- E &=& 0 \quad , \label{saa} \eea

\no For future purposes we note that if we use (\ref{a2a3}), the
equations (\ref{s11})-(\ref{s14}) become equivalent to

\be (S-1)(\ta - \tc) =0 \quad ; \quad (S-1)(\tb - \tc) =0 \quad .
\label{s1b} \ee

In order that equations (\ref{s01})-(\ref{s04}) have a nontrivial
solution for the pairs $(A,C)$ and $(B,C)$ a determinant must be
zero. It turns out that such determinant is exactly the constraint
(\ref{c6}). Likewise, a nontrivial solution for the equations
(\ref{s11})-(\ref{s14}) is warranted by the constraint (\ref{c4}).
Therefore, the absence of spin-0 and spin-1 poles in the
propagator requires the existence of both spin-0 and spin-1 local
symmetries in the corresponding massless theory. This is of course
analogous to the massive spin-1 Maxwell-Proca theory. The Maxwell
term, invariant under the usual spin-0 $U(1)$ gauge symmetry, is
singled out as the unique massless term by requiring the
non-propagation of the scalar mode $\p^{\mu} A_{\mu}$.

Regarding equation (\ref{saa}), since in the massless limit $d_- \to 0$, the constant $E$ is left arbitrary.
Consequently, the derivative terms of (\ref{lg}) are invariant under the following transverse antisymmetric
shifts (after redefining $\Lambda_{\mu\nu} \to \Box \Lambda_{\mu\nu}/E $):

\be \delta^{(0)}_{AA}e_{\alpha\beta} = [ P_{AA}^{(0)} ]_{\alpha\beta}^{\quad \mu\nu} \Box \Lambda_{\mu\nu} =
\p^{\mu}\Omega_{[\mu\alpha\beta ]} \equiv \lambda^{T}_{\alpha\beta} \quad , \label{aa} \ee

\no where $\p^{\alpha} \lambda^{T}_{\alpha\beta} = 0$ and

\be \Omega_{[\mu\alpha\beta ]} = \p_{\alpha} \Lambda_{[\mu\beta]} -  \p_{\beta} \Lambda_{[\mu\alpha]} - \p_{\mu}
\Lambda_{[\alpha\beta]}  \quad . \label{omega} \ee

\no The symmetry (\ref{aa}) is a consequence of the fact that the
derivative terms in (\ref{lgbh}) only depend upon $B_{\mu\nu}$
through the derivatives $\p^{\mu}B_{\mu\nu}$, see (\ref{lgbh}) and
(\ref{pmenos}).

It is clear from (\ref{c5}),(\ref{c6}),(\ref{c7}) and (\ref{s1b}) that $S=1$ and $S=2/3$ play a special role in
the parameters space of the general model (\ref{lg}). From now on we split our analysis into three classes,
namely, i) $S\ne 1 \, ; \, 2/3$ ; ii) $S=1$ ; iii) $S= 2/3$.

\subsection{$S\ne 1 \, ; \, 2/3$ }

In the massive case, if $S \ne 1$  we have from (\ref{s1b}) $\ta = \tb = \tc $. Back in (\ref{delta1}) the
spin-1 symmetry, after redefining $\Lambda \to \Box\Lambda /\tilde{A}$, becomes a transverse linearized
reparametrization\footnote{The role of this symmetry for describing massless spin-2 particles has been discussed
in \cite{bdng}, see also \cite{alvarez1}.}:

\be \delta^{(1)}e_{\mu\nu} =  [ P_{SS}^{(1)} +  P_{AA}^{(1)} +  P_{AS}^{(1)} + P_{SA}^{(1)} ]_{\mu\nu}^{\quad
\alpha\beta} \Box \Lambda_{\alpha\beta} = \p_{\nu}C_{\mu}^T \label{gt2} \ee

\no where $C^T_{\mu}=\p^{\beta}\left(\Lambda_{\mu\beta} - \Lambda_{\beta\mu}\right)$ satisfies
$\p^{\mu}C_{\mu}^T=0$.

If $S \ne 2/3$ we can obtain $b_1=b_1(b_2,S)$ from (\ref{c6}) and plug back in (\ref{s01})-(\ref{s04}) in order
to produce relationships between the constants $A,B,C$ such that we can, after some rearrangements, write down
the spin-0 symmetry:

\bea \delta^{(0)} e_{\mu\nu} &=& \left( x \, \theta_{\mu\nu} - 3\, y \, \omega_{\mu\nu} \right)\left( x
\,\theta^{\alpha\beta} - 3\, y \,\omega^{\alpha\beta} \right)\frac{\Lambda_{\alpha\beta}}{4(2/3 -S)} \nn \\
&=& (2\, S + b_2 - 1) \Box \Phi \eta_{\mu\nu} - 2 \, (S + 2\, b_2)\p_{\mu}\p_{\nu}\Phi \quad . \label{gt3} \eea

\no where the spin-0 and spin-1 projection operators $\omega_{\mu\nu}$ and $\theta_{\mu\nu}$ respectively are
defined in (\ref{pvectors}) and  $x=2\, S + b_2 - 1$ , $y=b_2 + 1/2$ and $ \Phi = \Box \left( x
\theta^{\alpha\beta} - 3\, y \omega^{\alpha\beta} \right)\Lambda_{\alpha\beta}/[4(2/3 -S)]$.

Therefore, if $S\ne 1$ and $S\ne 2/3$, the local symmetries of the massless theory become

\be \delta e_{\mu\nu} = \p_{\nu}C_{\mu}^T +  (2\, S + b_2 - 1) \Box \Phi \eta_{\mu\nu} - 2 \, (S + 2\,
b_2)\p_{\mu}\p_{\nu}\Phi + \lambda^{T}_{\mu\nu} \quad , \label{gt4} \ee

\no The transformation (\ref{gt4}) suggests us to split the
analysis further into two subcases  $S=-2\, b_2$ and $ S \ne -2\,
b_2$. If we plug $S=-2\, b_2$ back in the massive constraints
(\ref{c6}) and (\ref{c7}), recalling that $b_2 \ne -1/3$ due to $S
\ne 2/3$, we must have $c=m^2/8$ which violates our hypothesis
(\ref{d2}) and invalidates our particle content analysis in the
massive case. Indeed, if $S=-2\, b_2$,  with $b_2 \ne -1/2 \, ; \,
-1/3 $, it can be shown that the massive theory contains also a
scalar particle in the spectrum besides the massive spin-2 mode.
This is out of the scope of this work and $S=-2\, b_2$ will not be
considered here anymore except in the cases where $S=1 (b_2=-1/2)$
and $S=2/3 (b_2=-1/3)$ which are considered in the next
subsections.

If $ S \ne -2\, b_2$ we see from (\ref{w1}) that we can redefine $b_2$ as we wish. In particular, we can fix
$b_2 = 1- 2\, S$ which by the way holds for all models in the literature, see (\ref{fp}),(\ref{ms}) and
(\ref{nfp}). The symmetries (\ref{gt4}) of the massless theory become linearized reparametrizations plus
transverse antisymmetric shifts:

\be \delta e_{\mu\nu} = \p_{\nu}\xi_{\mu} + \lambda^T_{\mu\nu} \quad , {\rm if} \,\, b_2=1-2\, S  \quad .
\label{gt5} \ee

\no In order to figure out the particle content of the massless
theory we introduce an auxiliary vector field $C_{\mu}$ and
rewrite (\ref{lgbh}) in the massless limit (\ref{ml}) as

\bea {\cal L}_{m\to 0}(S) &=& (\p^{\mu}h_{\mu\nu})^2 + (1-2\, S)\p^{\alpha}h_{\alpha\mu}\p^{\mu}h + (\frac 12 -
S)\, h\Box h +
\frac 12 h_{\mu\nu}\Box h^{\mu\nu} \nn \\
&+& (1-S)\left\lbrack C^{\mu}C_{\mu} + 2 \, C^{\mu}
\left(\p^{\alpha}B_{\alpha\mu} + \p^{\alpha} h_{\alpha\mu}\right)
\right\rbrack \quad . \label{massless} \eea

\no If we perform the Gaussian integral over $C_{\mu}$ we recover our original model (\ref{lgbh}) in the
massless limit. If however, we first integrate over $B_{\mu\nu}$ in the path integral we have a functional
constraint assuring that $C_{\mu} = \p_{\mu} \phi $ for some scalar field $\phi$. Plugging back in
(\ref{massless}) and changing variables  $\phi = \varphi - h$ followed by $h_{\mu\nu} \to \tilde{h}_{\mu\nu} +
(1-S)\varphi\eta_{\mu\nu}$ we have the decoupled scalar-tensor theory:

\be {\cal L}_{m\to 0}(S) = {\cal L}_{LEH}[\tilde{h}_{\mu\nu}] - 3 \, (S-1)(S-2/3)\, \p^{\mu}\varphi \,
\p_{\mu}\varphi \quad . \label{ls} \ee

\no Where ${\cal L}_{LEH}$ is the usual linearized Einstein-Hilbert theory (or massless FP theory). Therefore,
we have a physical massless spin-2 particle plus a physical massless scalar particle as far as $S < 2/3 $  or $S
> 1$. Otherwise, the scalar particle becomes a ghost or disappears at $S=1$ or $S=2/3$. In
particular, the massive spin-2 model ($S=1/2$ and $b_2=0$) of \cite{ms} has a healthy scalar-tensor massless
limit. Regarding the symmetry (\ref{gt5}), since the equations of motion of (\ref{massless}) imply  $C_{\mu} =-
\p^{\alpha}e_{\alpha\mu} $ we deduce from (\ref{gt5}) and $C_{\mu}=\p_{\mu}\phi $ that $\delta \phi = - \p \cdot
\xi $. Consequently, $\delta \varphi = \delta (h + \Phi) = 0 $. Actually, this was the guideline for the
definition of $\varphi$ in the first place. Moreover, from $h_{\mu\nu} = \tilde{h}_{\mu\nu} +
(1-S)\varphi\eta_{\mu\nu}$ we have $\delta \tilde{h}_{\mu\nu} = \delta h_{\mu\nu} = \frac 12 (\p_{\mu}\xi_{\nu}
+ \p_{\nu}\xi_{\mu})$ which confirms the symmetry of (\ref{ls}) under (\ref{gt5}). In section 4 we return to the
$S\ne 1 \, ; \, 2/3$ family with nonzero mass.

\subsection{$S=1$}

If $S=1$, or equivalently $a_1 = 1/4$, we see from (\ref{c5}) that $d_-$ is a free parameter and from (\ref{c6})
and (\ref{c7}) we have respectively

\bea  b_1 &=& - (3\, b_2^2 + 2 \, b_2 + 1 )/4 \label{b1s1} \\
c &=& m^2 \, [1 + 3 (2 \, b_2 + 1)^2 ]/8 \label{cs1} \eea

\no Given our constraint $ c \ne m^2/8$, see (\ref{d1}), we see that $b_2 = -1/2$ plays a special role. This is
also expected from the fact that $S= - 2 \, b_2$  at this point.

We see from (\ref{s1b}) a peculiar feature of the $S=1$ case, namely, $\ta , \tb , \tc $ are arbitrary which
implies a maximal spin-1 symmetry in the massless limit, i.e., we have an independent symmetry generated by each
of the operators $P_{AA}^{(1)}$, $P_{SS}^{(1)}$ and $P_{AS}^{(1)} + P_{SA}^{(1)}$. We also have the spin-0
symmetries (\ref{aa}) and (\ref{gt3}) with $S=1$. In particular, from the symmetry generated by $P_{AA}^{(1)}$
and $P_{AA}^{(0)}$ and the antisymmetric closure relation (\ref{asym}), it is clear that we have symmetry under
arbitrary shifts in the antisymmetric sector ($\delta e_{\mu\nu} = \Lambda_{[\mu\nu ]}$) which allows us to get
rid of $e_{[\mu\nu]}$ in the $S=1$ case. Moreover, since the transformations generated by $P_{AS}^{(1)}$ lie in
the antisymmetric sector, which can be gauged away, we just need to worry about $P_{SS}^{(1)}$ and
$P_{SA}^{(1)}$ which give rise to

\be \delta_{SS}^{(1)} e_{\mu\nu} + \delta_{SA}^{(1)} e_{\mu\nu} = \p_{\mu}C_{\nu}^T + \p_{\nu}C_{\mu}^T
\label{ss1} \ee

\no where the transverse vector is now given by

\be C_{\mu}^T = \Box \p^{\alpha} ( \Lambda_{\alpha\mu} + \Lambda_{\mu\alpha} ) - 2
\p_{\mu}\left(\p^{\alpha}\p^{\beta}\Lambda_{\alpha\beta}\right) + \p^{\alpha}( \Lambda_{\alpha\mu} -
\Lambda_{\mu\alpha} ) \quad . \label{ct01} \ee

\no Altogether from (\ref{gt3}) and (\ref{ss1}) we have the whole set of local symmetries of the massless $S=1$
case given by

\be \delta e_{\mu\nu} = \p_{\mu}\xi_{\nu} + \p_{\nu}\xi_{\mu} + (1 + b_2) \eta_{\mu\nu} \, \Box \, \Phi +
\Lambda_{[\mu\nu]} \quad , \label{gt6} \ee

\no where

\be \xi_{\mu} = C_{\mu}^T - (1 + 2\, b_2) \p_{\mu} \Phi \quad . \label{xi} \ee

\no If $b_2 = -1/2$ ($S=-2\, b_2$) we can redefine $\Phi \to 2 \, \Phi/\Box $ and write down the symmetry of the
massless theory

\be \delta e_{\mu\nu} = \p_{\mu} C_{\nu}^T + \p_{\nu} C_{\mu}^T + \eta_{\mu\nu}\, \Phi + \Lambda_{[\mu\nu]}
\quad , \quad {\rm if} \, b_2=-1/2 \quad . \label{gt7} \ee

\no The above case is known as WTDIFF theory, see \cite{abgv}, due
to the Weyl symmetry and transverse (linearized) diffeomorphisms.
It is the only possible description of a massless spin-2 particle
in terms of one symmetric rank-2 tensor which differs from the
usual massless Fierz-Pauli theory (linearized Einstein-Hilbert).
It admits a nonlinear extension known as unimodular gravity
\cite{abgv}. If we add a mass term, the theory becomes unstable
\cite{abgv}.

If $b_2 \ne -1/2$ ($S \ne - 2 \, b_2$) we can bring $b_2 \to -1$ ($b_2 = 1-2\,S$) and end up with the usual
massless Fierz-Pauli theory describing a $m=0$ spin-2 particle which admits the usual massive extension with
mass term $m^2(e_{\mu\nu}e^{\nu\mu} - e^2)$. The symmetries of the massless theory (\ref{gt6})  become the usual
linearized diffeomorphisms plus arbitrary antisymmetric shifts $ \delta e_{\mu\nu} = \p_{\mu}\xi_{\nu} +
\p_{\nu}\xi_{\mu}  + \Lambda_{[\mu\nu]} $.

\subsection{ $S=2/3$ }

In this case we have $a_1 = - 1/12$ and from (\ref{c6}) and
(\ref{c7}) we have $b_2 = -1/3$ and $b_1=-1/6$ while $c$ is a free
parameter. The peculiar feature of this case is the maximal set of
spin-0 symmetries in the massless theory with $A,B,C$ arbitrary,
see (\ref{s01})-(\ref{s04}). Since $d_- \to 0$ in the massless
limit, each of the spin-0 operators
$P_{AA}^{(0)},P_{WW}^{(0)},P_{SS}^{(0)},P_{SW}^{(0)}+
P_{WS}^{(0)}$ generates a symmetry. Altogether it can be shown
that those spin-0 symmetries can be written as

\be \delta e_{\mu\nu} = \p_{\nu} \xi_{\mu} + \eta_{\mu\nu} \phi + \Lambda_{\mu\nu}^T \phi \ee

\no The massless $S=2/3$ theory has been first analyzed in
\cite{cmu} and later in \cite{nfp}. It describes one massless
spin-2 particle in terms of a nonsymmetric tensor. The massive
case is studied in \cite{nfp} and describes one massive spin-2
particle. We can extend the Weyl symmetry to the massive case for
the choice $c=m^2/8$. For more details see \cite{cmu,nfp}.

\section{ A new family of massive spin-2 models}

In the last section we have seen that all one-parameter models
describing one massive spin-2 particle out of a rank-2 tensor can
be classified in three classes. Given that two of them ($S=1$ and
$S=2/3$) have already appeared in the literature, we now focus on
the new family of models defined in terms of the free parameter
$a_1$ which we call\footnote{Alternatively we could use the sum
$S=a_1+3/4$ as free parameter.} ${\cal L}(a_1)$. The new family
corresponds to the coefficients :

\bea d_1 = 0 \, ; \, d_2=-m^2/2 \, &;& \, a_2=1/2 \quad ; \quad a_3 = 1/4  \, ,  \nn \\
 b_1 = -(a_1 + 1/4) \, &;& \, b_2 = - 2(a_1 + 1/4) \quad ; \quad c=m^2/2 . \label{nf} \eea

\no Explicitly, from (\ref{lg}) we have

\bea {\cal L}(a_1) &=& a_1
\left(\p^{\alpha}e_{\alpha\beta}\right)^2 + \frac 12
\left(\p^{\alpha}e_{\alpha\nu}\right)\left(\p_{\beta}e^{\nu\beta}\right)
+ \frac 14 \left(\p^{\nu}e_{\mu\nu}\right)^2 + \left(a_1 + \frac
14\right) \p^{\mu}e\, (\p_{\mu}e - 2 \, \p^{\alpha}e_{\alpha\mu})
 \nn \\
&+& \frac 14 e_{\mu\nu} \Box ( e^{\mu\nu} +  e^{\nu\mu}) -
\frac{m^2}2 \left( e_{\mu\nu}e^{\nu\mu} - e^2 \right) \quad .
\label{la1} \eea

\no The equations of motion of (\ref{la1}) become

\bea \Box \, e^{(\mu\nu)} &-& (2\, a_1 + 1/2)\left\lbrack \eta^{\mu\nu}(\Box \, e - \p^{\alpha}\p^{\beta}
e_{\alpha\beta} ) -
\p^{\mu}\p^{\nu}e \right\rbrack \nn\\
&=& \p^{(\mu} \p^{\alpha} e^{\nu)}_{\,\, \alpha} + 2\, a_1
\p^{\mu}\p^{\alpha}e_{\alpha}^{\,\,\nu} + \frac 12
\p^{\nu}\p^{\alpha}e_{\alpha}^{\,\, \mu} -m^2 (\eta^{\mu\nu}\, e -
e^{\nu\mu}) \quad . \label{ema1} \eea

\no Applying $\p_{\nu}$ on (\ref{ema1}) we derive the constraint

\be \p_{\nu}e^{\nu\mu}=\p^{\mu} e \quad . \label{t1} \ee

\no By plugging back (\ref{t1}) in (\ref{ema1}) the antisymmetric part of the resulting equation leads to
another constraint:

\be e_{[\mu\nu]} = 0 \quad . \label{t2} \ee

\no From the trace of (\ref{ema1}), using (\ref{t1}) and
(\ref{t2}), we derive

\be e =0 \quad. \label{t3} \ee

\no Therefore, back in (\ref{t1}) we have the transversality relation:

\be \p_{\mu}e^{\mu\nu} = 0 = \p_{\nu}e^{\mu\nu} \quad . \label{t4} \ee

\no Finally, (\ref{ema1}) becomes the Klein-Gordon equation:

\be (\Box - m^2)e^{(\mu\nu)} = 0 \quad . \label{kg} \ee

\no In conclusion we have a massive spin-2 particle with the
correct counting of degrees of freedom for arbitrary values of
$a_1$. If $a_1=1/4$ ($S=1$) we recover the massive FP theory
 while $a_1=-1/12$ ($S=2/3$) and $a_1=-1/4$ ($S=1/2$) lead to the
other two models of the literature which describe massive spin-2
particles via a rank-2 tensor, \cite{nfp} and \cite{ms}
respectively, . Thus, the one-parameter family ${\cal L}(a_1)$
intersects the other two classes $S=1$ ($b_2 \ne -1/2$) and
$S=2/3$ at the specific points where the corresponding free
parameters of those classes become respectively $d_- = m^2/2$ and
$c=m^2/2$. So ${\cal L}(a_1)$ contains all known models for a
massive spin-2 particle.

We finish this section commenting on the vDVZ mass discontinuity
for the ${\cal L}(a_1)$ family. Since in the massive case the only
singular term of the propagator is of the same form of the massive
FP theory ($a_1$-independent). Disregarding terms which are not
important for the light beams deviation by the sun we have for the
massless limit of the massive propagator

\be \left(G_{a_1}^{-1}\right)^{{\rm sing}}_{\mu\nu\alpha\beta} =
\lim_{m\to 0} \frac 2{\Box - m^2}
\left(P_{SS}^{(2)}\right)_{\mu\nu\alpha\beta} = \frac
2{\Box}\left( \frac{\eta_{\mu\alpha}\eta_{\nu\beta} +
\eta_{\mu\beta}\eta_{\nu\alpha}}2 -
\frac{\eta_{\mu\nu}\eta_{\alpha\beta}}3 \right) + \cdots \, ,
\label{ga1m} \ee

\no where dots (here and in the next formula) stand for
unimportant terms for the light beams deviation. The discrepancy
of the deviation angle will be the same one of the massive FP
theory $\theta(a_1) = \theta(FP) = (3/4)\theta_{GR} $.

On the other hand, for the massless model ${\cal L}_{m \to
0}(a_1)$ the relevant piece of the propagator is given by

\bea \left(G_{a_1}\right)^{{\rm sing}}_{\mu\nu\alpha\beta} &=&
\left\lbrack \frac 2{\Box } P_{SS}^{(2)} - \frac{4 \,
P_{SS}^{(0)}}{\Box \, (1 + 12 \, a_1)} + \cdots
\right\rbrack_{\mu\nu\alpha\beta} \nn\\ &=& \frac 2{\Box}
\left\lbrack \frac{\eta_{\mu\alpha}\eta_{\nu\beta} +
\eta_{\mu\beta}\eta_{\nu\alpha}}2 - \frac{(1 + 4 \, a_1)}{(1+12\,
a_1)} \eta_{\mu\nu}\eta_{\alpha\beta} \right\rbrack + \cdots
\label{ga10} \eea

\no Therefore, we have the deviation angle

\be \theta_{m=0} (a_1) = \left( \frac{16 \, a_1}{1+12\,
a_1}\right) \theta_{GR} \quad . \label{tgr} \ee

\no Thus, the massless theory only reproduces the GR result in the
trivial case $a_1=1/4$ where ${\cal L}(a_1) = {\cal L}_{LEH}$.
However, since $\theta_{m=0} (a_1)$ continuously approaches
$\theta_{GR}$ from above as $a_1 \to (1/4)^+$ we may have $a_1$
close enough to $1/4$ such that the difference $\theta_{m=0}
(a_1)-\theta_{GR}$ is below the experimental error bar. So we can
not discard the massless scalar-tensor theory ${\cal
L}_{m=0}(a_1)$ based on the light beams deviation by the sun.
Recall that $S=a_1 + 3/4$, consequently the ghost-free bounds $S
\ge 1$ or $S \le 2/3$ contain the region where $a_1$ is slightly
above $1/4$.

\section{Conclusion}

We have started with a second-order theory  for a general rank-2
tensor with 10 free parameters. Requiring that we only have one
massive spin-2 particle in the spectrum we are left with only one
free parameter up to the local field redefinitions (\ref{w}) and
(\ref{r}). We have proved that in the massless limit we always
have one spin-0 plus one spin-1 local symmetry. The use of spin
projection and transition operators and their algebra, see
appendix A, has helped us in splitting the one-parameter family of
models into three classes: $S=1$, $S=2/3$, and $S \ne 1 \, ; \,
2/3$, where $S=a_1+a_2+a_3$ is invariant under the field
redefinitions (\ref{w}) and (\ref{r}). In the massless limit, if
$S=1$ the local spin-1 symmetry is maximal while $S=2/3$
corresponds to a maximal spin-0 symmetry.

In the class $S=1$ the coefficient $d_-$ in (\ref{lg}) is the free
parameter. We can get rid of the antisymmetric part of the tensor
via a local gauge symmetry and work with a purely symmetric
tensor. If $S=1$ and $b_2 \ne -1/2$ the one-parameter family of
models is equivalent, after a field redefinition, to the well
known massive Fierz-Pauli theory, thus describing a massive spin-2
particle. The special case $S=1$ and $b_2=-1/2$ is unstable
\cite{abgv}. Its massless limit is invariant under linearized Weyl
and linearized transverse reparametrizations. It is the WTDIFF
model of \cite{abgv}. It describes a massless spin-2 particle and
it is the linearized version of unimodular gravity. We remark that
here we never have the so called TDIFF theories which are
invariant under transverse linearized reparametrizations. This
follows from our primary assumption that our massive theory only
describes a physical massive spin-2 particle while in  TDIFF
theories there is always a scalar particle \cite{abgv}.

In the class $S=2/3$ the coefficient $c$ in (\ref{lg}) becomes a
free parameter. The derivative terms are invariant under a
linearized Weyl transformation such that the trace
$e=\eta^{\mu\nu}e_{\mu\nu}$ only appears non dynamically in the
mass term. Although the trace is nonzero off shell, as far as we
remains at the quadratic level (free theory) it decouples and
plays no role. So effectively we have a traceless (though
nonsymmetric) description of a massive spin-2 particle \cite{nfp}.
In the massless limit we have one massless spin-2 particle, see
\cite{cmu} and also \cite{nfp}.

In the third class of models, $S \ne 1 \, ; \, 2/3$, the parameter $S$ itself becomes the free parameter. There
is no further restriction on $S$ in the massive theory. It does represent a new one-parameter family of models
describing a massive spin-2 particle. At the point $S=1/2$ we recover the model of \cite{ms}. The massive theory
has the same mass discontinuity problem of the massive Fierz-Pauli theory and predicts the same wrong deviation
angle of the light beams by the sun which is independent of $S$. In the massless limit we have a scalar-tensor
theory whose unitarity requires $ S \le 2/3$ or $S \ge 1$. For the scalar-tensor theory the deviation angle can
be made consistent with experimental data if $S$ is chosen slightly above one. General relativity is recovered
at $S=1$.

The one-parameter families $S=2/3$ and $S \ne 1 \, ; \, 2/3$ are ghost- and tachyon-free although we have a
coupling between the antisymmetric and symmetric parts of the rank-2 tensor. This is contrary to the claim of
\cite{pvn73} that this kind of coupling will necessarily lead to ghosts. In the specific example of
antisymmetric/symmetric coupling chosen in \cite{pvn73} there is in fact a ghost but it is not the general
situation as shown here and in the earlier examples \cite{cmu} (massless case) and \cite{ms,nfp} (massive
cases).

Finally, an arbitrary tensor $e_{\mu\nu}$ can be decomposed into a
traceless symmetric field ($h_{\mu\nu}^T$), a pure trace piece ($h
\, \eta_{\mu\nu}$)  plus an antisymmetric tensor ($B_{\mu\nu}$).
Since a massive spin-2 particle requires on shell that
$h=0=B_{\mu\nu}$, the fields $h$ and $B_{\mu\nu}$ are auxiliary.
It is well known, see for instance \cite{barnes}, that we can not
set $h=0=B_{\mu\nu}$ off shell. So the next minimal possibility is
to set only $B_{\mu\nu}=0$ off shell, this is indeed possible and
corresponds to the usual massive Fierz-Pauli theory or linearized
Einstein-Hilbert theory. In our notation it corresponds to the
$S=1$ case with $b_2 \ne -1/2$. It has been shown in \cite{nfp}
that the next simplest case with $h=0$ and $B_{\mu\nu} \ne 0$,
both off shell, is also possible. It is the $S=2/3$ case. Here we
have shown that we are allowed to keep both auxiliary fields $h$
and $B_{\mu\nu}$ non vanishing off-shell ($S \ne 1 ; 2/3$ case)
without any inconsistency as far as we deal with the free theory.
Since auxiliary fields may become dynamical when we turn on
interactions and lead to troubles like incorrect counting of
degrees of freedom (loss of constraints), ghosts and acausality,
see e.g. \cite{vz,db} and more recently \cite{dgnr}-\cite{dw}, it
is crucial to  investigate the addition of interactions to the
massive spin-2 models with $S \ne 1$. This is now in progress.

\section{Appendix A}

After defining the spin-0 and spin-1 projection operators, respectively,

\be  \omega_{\mu\nu} = \frac{\p_{\mu}\p_{\nu}}{\Box} \quad , \quad   \theta_{\mu\nu} = \eta_{\mu\nu} -
\frac{\p_{\mu}\p_{\nu}}{\Box}\quad , \label{pvectors} \ee

\no one can define the projection and transition operators, see e.g.  \cite{pvn73}.  First we present the
symmetric operators

\be \left( P_{TT}^{(2)} \right)^{\lambda\mu}_{\s\s\alpha\beta} = \frac 12 \left(
\theta_{\s\alpha}^{\lambda}\theta^{\mu}_{\s\beta} + \theta_{\s\alpha}^{\mu}\theta^{\lambda}_{\s\beta} \right) -
\frac{\theta^{\lambda\mu} \theta_{\alpha\beta}}{3} \quad , \label{ps2} \ee

\be \left( P_{SS}^{(1)} \right)^{\lambda\mu}_{\s\s\alpha\beta} = \frac 12 \left(
\theta_{\s\alpha}^{\lambda}\,\omega^{\mu}_{\s\beta} + \theta_{\s\alpha}^{\mu}\,\omega^{\lambda}_{\s\beta} +
\theta_{\s\beta}^{\lambda}\,\omega^{\mu}_{\s\alpha} + \theta_{\s\beta}^{\mu}\,\omega^{\lambda}_{\s\alpha}
 \right) \quad , \label{ps1} \ee

\be \left( P_{TT}^{(0)} \right)^{\lambda\mu}_{\s\s\alpha\beta} = \frac 1{3} \,
\theta^{\lambda\mu}\theta_{\alpha\beta} \quad , \quad \left( P_{WW}^{(0)} \right)^{\lambda\mu}_{\s\s\alpha\beta}
= \omega^{\lambda\mu}\omega_{\alpha\beta} \quad , \label{psspww} \ee

\be \left( P_{TW}^{(0)} \right)^{\lambda\mu}_{\s\s\alpha\beta} = \frac 1{\sqrt{3}}\,
\theta^{\lambda\mu}\omega_{\alpha\beta} \quad , \quad  \left( P_{WT}^{(0)}
\right)^{\lambda\mu}_{\s\s\alpha\beta} = \frac 1{\sqrt{3}}\, \omega^{\lambda\mu}\theta_{\alpha\beta} \quad ,
\label{pswpws} \ee

\no They satisfy the symmetric closure relation

\be \left\lbrack P_{TT}^{(2)} + P_{SS}^{(1)} +  P_{TT}^{(0)} + P_{WW}^{(0)} \right\rbrack_{\mu\nu\alpha\beta} =
\frac{\eta_{\mu\alpha}\eta_{\nu\beta} + \eta_{\mu\beta}\eta_{\nu\alpha}}2 \quad . \label{sym} \ee

\no The remaining antisymmetric and mixed symmetric-antisymmetric operators are given by

\be \left( P_{AA}^{(1)} \right)^{\lambda\mu}_{\s\s\alpha\beta} = \frac 12 \left(
\theta_{\s\alpha}^{\lambda}\,\omega^{\mu}_{\s\beta} - \theta_{\s\alpha}^{\mu}\,\omega^{\lambda}_{\s\beta} -
\theta_{\s\beta}^{\lambda}\,\omega^{\mu}_{\s\alpha} + \theta_{\s\beta}^{\mu}\,\omega^{\lambda}_{\s\alpha}
 \right) \quad , \label{paa1} \ee

\be \left( P_{SA}^{(1)} \right)^{\lambda\mu}_{\s\s\alpha\beta} = \frac 12 \left(
\theta_{\s\alpha}^{\lambda}\,\omega^{\mu}_{\s\beta} + \theta_{\s\alpha}^{\mu}\,\omega^{\lambda}_{\s\beta} -
\theta_{\s\beta}^{\lambda}\,\omega^{\mu}_{\s\alpha} - \theta_{\s\beta}^{\mu}\,\omega^{\lambda}_{\s\alpha}
 \right) \quad , \label{pas1} \ee

\be \left( P_{AS}^{(1)} \right)^{\lambda\mu}_{\s\s\alpha\beta} = \frac 12 \left(
\theta_{\s\alpha}^{\lambda}\,\omega^{\mu}_{\s\beta} - \theta_{\s\alpha}^{\mu}\,\omega^{\lambda}_{\s\beta} +
\theta_{\s\beta}^{\lambda}\,\omega^{\mu}_{\s\alpha} - \theta_{\s\beta}^{\mu}\,\omega^{\lambda}_{\s\alpha}
 \right) \quad , \label{psa1} \ee

\be \left( P_{AA}^{(0)} \right)^{\lambda\mu}_{\s\s\alpha\beta} = \frac 12 \left(
\theta_{\s\alpha}^{\lambda}\theta^{\mu}_{\s\beta} - \theta_{\s\alpha}^{\mu}\theta^{\lambda}_{\s\beta} \right)
\quad , \label{paa0} \ee

\no They satisfy the antisymmetric closure relation

\be \left\lbrack P_{AA}^{(1)} + P_{AA}^{(0)} \right\rbrack_{\mu\nu\alpha\beta} =
\frac{\eta_{\mu\alpha}\eta_{\nu\beta} - \eta_{\mu\beta}\eta_{\nu\alpha}}2 \quad . \label{asym} \ee

\no Adding up (\ref{sym}) and (\ref{asym}) we have

\be \left\lbrack P_{TT}^{(2)} + P_{SS}^{(1)} +  P_{TT}^{(0)} + P_{WW}^{(0)} +  P_{AA}^{(1)} + P_{AA}^{(0)}
\right\rbrack_{\mu\nu\alpha\beta} = \eta_{\mu\alpha}\eta_{\nu\beta} \quad . \label{full} \ee

\section{Appendix B}

Here we prove, with help of the field redefinition (\ref{r}), that we can always choose $(a_2,a_3)=(1/2,1/4)$ as
a solution of (\ref{c4}) without loss of generality.

It is clear from (\ref{r1})-(\ref{r3}) that the combination $a_1 +
a_3 - a_2 $ plays a special role. It is convenient to rewrite
(\ref{c4}) as

\be (S-1)(a_1 + a_3 - a_2) = (a_1-a_3)^2 \quad . \label{c4b} \ee

\no If $a_1 + a_3 - a_2 = 0$ we must have  $a_1 = a_3$. Back in
(\ref{c5}) we have, since (\ref{d1}) demands $d_- \ne 0$, that
$S=a_1+a_2+a_3 = 1 $. Those equations fix $a_2=1/2$ and $a_1 = a_3
= 1/4$. Regarding the hypothesis $d_- \ne 0$ note that, since $a_1
+ a_3 - a_2 = 0 = a_1 - a_3 $, the antisymmetric tensor
$B_{\mu\nu}$ only appears in the Lagrangian (\ref{lgbh}) through
the trivial term $d_-B_{\mu\nu}B^{\mu\nu}$. Therefore there will
be no change in the physical content of the theory if we assume
$d_- \ne 0$. We conclude that if $a_1 + a_3 - a_2 = 0$ we
automatically have $(a_2,a_3)=(1/2,1/4)$.

On the other hand, if $a_1 + a_3 - a_2 \ne 0$ we can always move $a_2$ from any given value to $a_2=1/2$ via
(\ref{r1}) by choosing $A=(2\, a_1 - a_2)/[2(a_1 + a_3 - a_2)]$ and using (\ref{c4}). Back in (\ref{r3}) we
have, using again (\ref{c4}), the new value $a_3 = (4\, a_1 a_3 - a_2^2)/[4(a_1 + a_3 - a_2)]=1/4$. This
completes the proof of (\ref{a2a3}).

\section{Acknowledgements}

This work is supported by CNPq and FAPESP (2013/00653-4).

\end{document}